# Three-dimensional bi-functional refractive index and fluorescence microscopy (BRIEF)

Yi Xue[1], David Ren[1], Laura Waller[1,*]

[1]Department of Electrical Engineering & Computer Sciences, University of California, Berkeley, CA, USA

[*]Corresponding author: waller@berkeley.edu

**Abstract**

Fluorescence microscopy is a powerful tool for imaging biological samples with molecular specificity. In contrast, phase microscopy provides label-free measurement of the sample's refractive index (RI), which is an intrinsic optical property that quantitatively relates to cell morphology, mass, and stiffness. Conventional imaging techniques measure either the labeled fluorescence (functional) information *or* the label-free RI (structural) information, though it may be valuable to have both. For example, biological tissues have heterogeneous RI distributions, causing sample-induced scattering that degrades the fluorescence image quality. When both fluorescence and 3D RI are measured, one can use the RI information to digitally correct multiple-scattering effects in the fluorescence image. Here, we develop a new computational multi-modal imaging method based on epi-mode microscopy that reconstructs both 3D fluorescence and 3D RI from a single dataset. We acquire dozens of fluorescence images, each 'illuminated' by a single fluorophore, then solve an inverse problem with a multiple-scattering forward model. We experimentally demonstrate our method for epi-mode 3D RI imaging and digital correction of multiple-scattering effects in fluorescence images.

**Main**

Fluorescence microscopy and phase microscopy are two distinct imaging techniques that leverage different contrast mechanisms: fluorescence microscopy images specific structures that are labeled by fluorescent tags in a biological sample; phase microscopy, on the other hand, images the refractive index (RI) of a sample and can be used for visualizing label-free structures, while lacking molecular specificity. Multimodal microscopy methods that can image both fluorescence-labeled and label-free structures enables correlating the two. The phase images reconstruct the RI of the sample, which may give structural information about the sample, or can be used to correct sample-induced scattering effects computationally in the fluorescence.

Previous work either focused on reconstructing fluorescence signals through tissue scattering by wavefront shaping[1-3], ultrasound-assisted optical imaging[4-7], measurement of transmission[8,9] or reflection matrix[10,11], and computational optimization[12-16]; or focused on measuring RI with optical diffraction tomography[17-19], computational phase retrieval[20-24], or optical coherence refractive tomography[25]. A few existing methods recover both fluorescence and phase information[26-29], but they require sequential experiments or independent measurements, which makes it difficult to register the images spatially. No previous method reconstructs both RI and fluorescence from a single dataset captured by one camera. One of the reasons is that fluorescence microscopy is usually in epi-mode, while phase microscopy is usually in transmission mode, which limits the application for *in vivo* imaging. Here, we introduce an epi-mode multimodal microscopy system that merges the function of fluorescence microscopy and epi-mode ODT, which could become a powerful tool for bioimaging.

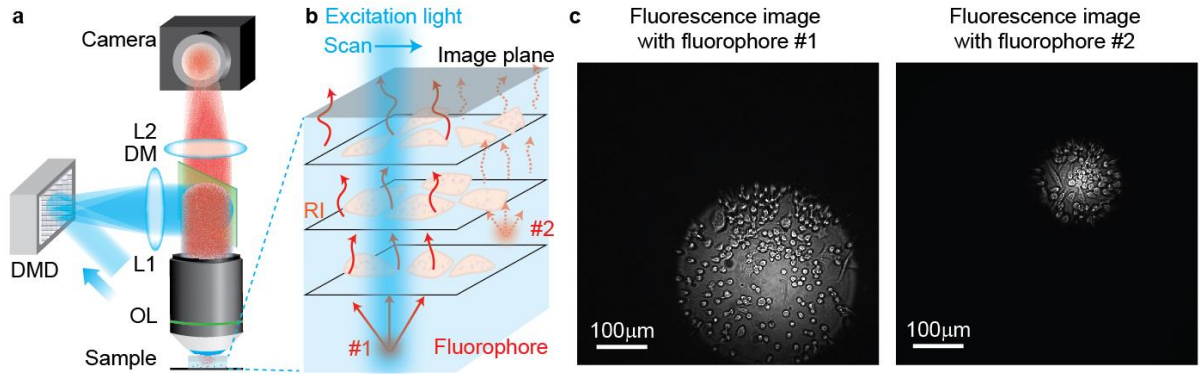

**Fig. 1: 3D bi-functional refractive index and fluorescence microscopy (BRIEF).** (a) In order to excite different fluorophores at different times, a collimated illumination beam is modulated by a digital micromirror device (DMD) to selectively excite each fluorophore sequentially. Emitted fluorescence light scatters through the label-free tissue above and is imaged onto a camera sensor, which is focused at the top of the sample. (b) The sample here consists of fluorescence-labeled structures on the bottom and non-labeled cells on the top. A multi-slice scattering algorithm models how fluorescence light propagates through each depth slice of the sample, scattering according to its 3D refractive index (RI). (c) Two examples of raw images captured by the camera, corresponding to fluorophore #1 and #2 in (b) being turned 'on'. The size of the circular region is determined by the depth of the fluorophore and the numerical aperture (NA) of the system; fluorophore #1 is located deeper than fluorophore #2, so it illuminates a larger volume. Fine structures inside of the circular area phase information about the cells and can be considered as intensity images taken with different defocus and/or illumination angles.

We provide a proof-of-principle demonstration for 3D bi-functional refractive index and fluorescence microscopy (BRIEF), which reconstructs both 3D fluorescence and 3D RI from fluorescence images captured in epi-mode. We focus the microscope at the top of the sample and collect dozens of images, each having different fluorophores within the volume 'on'. The fluorophores deeper in the sample illuminate different parts of the phase objects near the surface from different angles. We can estimate the fluorophore 3D position from the fluorescence images and also reconstruct the 3D RI, as long as the set of captured images contains diverse illumination angles for each lateral position. To recover the 3D information, we solve an inverse problem with a physics-based multi-slice model that accounts for multiple scattering. Because both modalities are reconstructed from the same dataset captured by a single camera, the fluorescence and RI signals are strictly registered in space and time.

The experimental setup of BRIEF is shown in Fig. 1. Our test samples are 3D phase objects (beads or cells) seeded with fluorophores that are sufficiently deep in the sample (hundreds of microns) to illuminate the phase objects from below. In general, BRIEF could be used to image fluorescence-labeled biological tissue which has either sparsely blinking fluorophores in scattering tissue (e.g. the dyes used in super-resolution localization microscopy[30, 31], or the fluorophores can be selectively illuminated by a photo-stimulation setup. For the latter case, one can use a widefield fluorescence stack to identify fluorophores' locations first, followed by stimulating them sequentially; or, one can scan an illumination 'point source' through the sample volume and capture a 2D image at each scanning location, then only use the images in which a fluorophore is emitting. We choose the latter of these approaches for convenience. We modulate collimated laser light at 473nm wavelength with a Digital Micromirror Device (DMD) located at the relayed image plane (Fig. 1a). The pixels on the DMD are binned into 20x20 pixel patches (corresponding to 8x8µm$^2$ at the sample), termed "super-pixels" and each super-pixel is turned on one-by-one to scan the volume laterally. Given weak scattering and sparse fluorophore distribution, either a single fluorophore or no fluorophore is excited at each scanning location; we select the images containing fluorophores as our measurements (Fig. 1c). The fluorophores act like point light sources inside the sample, creating a circular bright area in the measurement, inside which we see fine structures due to the phase objects (live CHO cells). The size of the illuminated area is proportional to the depth of the fluorophore: deeper fluorophores will give larger illuminated areas. Fluorophores located at different positions will illuminate different parts of the phase object from different angles. We collect dozens fluorescence images, each illuminated by a different fluorophore, as raw measurements, such that each part of the phase object sees a diversity of illumination angles over the set of captured images.

The 3D information in BRIEF comes from illuminating the phase objects from different angles via the spherical waves generated by the fluorophores at various depths. This is somewhat analogous to a fan-beam version of ODT, which illuminates phase objects from different angles in transmission mode. Therefore, BRIEF can incorporate similar forward models as ODT for 3D RI reconstruction. Here, we use one with a multi-slice scattering model[21, 23, 32] and treat the fluorescence sources inside the tissue, denoted by $o_f(\boldsymbol{r})$, as spherical emitters, where each fluorophore's light is spatially coherent (with itself) but incoherent with other fluorophores. The multi-slice model approximates the bulk tissue as a series of thin layers, where the RI of the $k$th layer is denoted by $n_k(\boldsymbol{r})$. Light propagation through the tissue is modeled via sequential layer-to-layer propagation of the electric field. The intensity of the exit electric field at the image plane $z$, $I_l(\boldsymbol{r}; z)$, accounting for the accumulation of diffraction and multiple scattering, is recorded by a sensor as the $l$th measurement. Both the 3D fluorescence distribution, $o_f(\boldsymbol{r})$, and the 3D RI, $n(\boldsymbol{r})$, are unknown in the model. However, the approximate position of fluorescence objects can be estimated from the measurements if the scattering is weak and isotropic, such that the intensity image has a clear circle area illuminated. In the lateral direction, the fluorophore position is approximately the center of gravity of the circle, which is the initial value of $o_f(\boldsymbol{r})$. In the axial direction, the fluorophore position is estimated by fitting to the point-spread-function (PSF) at each axial depth. Therefore, we first estimate the 3D fluorescence distribution $o_f(\boldsymbol{r})$ from multiple 2D fluorescence images and then use the expected fluorescence positions to estimate the 3D RI, $n(\mathbf{r})$, by solving an optimization problem with Tikhonov regularization, denoted by $R$:

$$\underset{n(\mathbf{r})}{\operatorname{argmin}} \sum_l \left\| I_l(\mathbf{r}; z) - \hat{I}_l(\mathbf{r}; z) \right\|_2^2 + R[n(\mathbf{r})]. \qquad (1)$$

We solve this optimization problem with the fast iterative shrinkage-thresholding algorithm (FISTA), which is a first-order gradient descent algorithm. Details about the forward model and reconstruction process are in the Methods section.

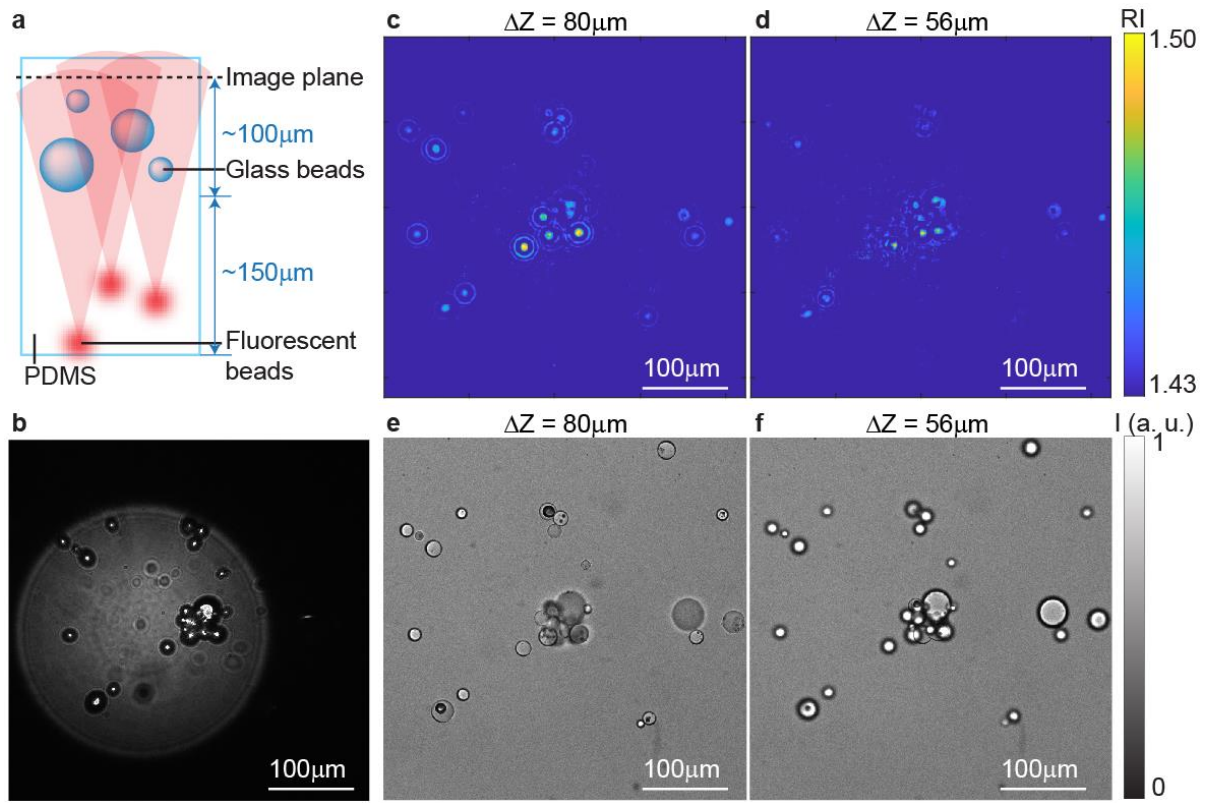

Fig. 2. BRIEF for reconstructing 3D RI of glass beads from a single dataset of experimentally measured fluorescence images taken with different fluorescent beads emitting. **a**. The sample consists of fluorescent beads on the bottom layer of the PDMS to act as light sources and glass beads on the top layer of the PDMS to act as non-fluorescence phase objects to be reconstructed. **b**. One of the 21 raw measurements used for reconstruction, with the defocused fluorescence signal scattered by the glass beads. **c, d**. Reconstructed RI of glass beads at $\Delta z = 80 \mu m$ and $\Delta z = 56 \mu m$ below the image plane, respectively. **e, f**. Widefield images of the sample under transmitted illumination to show the position of the glass beads as a 'ground truth' to compare with the reconstruction results.

We first demonstrate that our method can reconstruct 3D RI from fluorescence images by using a calibrated sample with ground truth information of its 3D RI. The calibration sample consists of two layers: the bottom layer of PDMS (RI 1.43) is about 150 μm thick with 0.71μm red fluorescence beads inside as the light sources (positions are unknown); the top layer of PDMS is about 100μm thick with glass beads of known size and RI (RI 1.50, size 5-50μm) inside as weak scattering phase objects (Fig. 2a). To find the positions of the glass beads for our ground truth information, we capture transmission widefield images at several axial planes with infrared LED illumination (Fig. 2e-f). For our BRIEF reconstruction, we capture 21 fluorescence measurements with different fluorophores on. A representative measurement is shown in Fig. 2b, where we can see a defocused fluorescence light source illuminating some of the glass beads. After our algorithm, the reconstructed 3D RI clearly distinguishes each glass bead in a cluster with good optical sectioning ability (Fig. 2c-d), and the 3D position of the glass beads matches with the ground truth widefield focus stack. This experimental result demonstrates our ability to reconstruct 3D RI from fluorescence images.

Next, we demonstrate BRIEF for reconstructing both fluorescence and RI from the same raw measurements of live biological cells. The test sample in this case is made by two steps: first, we fixed red fluorescent beads in PDMS on top of a coverslip; next, we coated the top surface of the PDMS with poly-lysine and cultured a thin layer of non-labeled CHO cells on it, in order to obtain a realistic biological phase object. During the experiment, we placed the sample in phosphate-buffered saline solution (RI 1.33) and collected 23 fluorescence images. An example of the fluorescence images for one excited fluorescent bead is shown in Fig. 3b. We treat the previous experiment of glass beads as a calibration reference and used the same parameters in the optimization algorithm for reconstruction. The reconstructed RI of CHO cells at $\Delta z = 40 \mu m$ below the image plane is shown in Fig. 3c, and Fig. 3d shows the position of the cells using the transmission widefield image with infrared LED illumination (note that these are different contrast measurements so the images should not be compared directly).

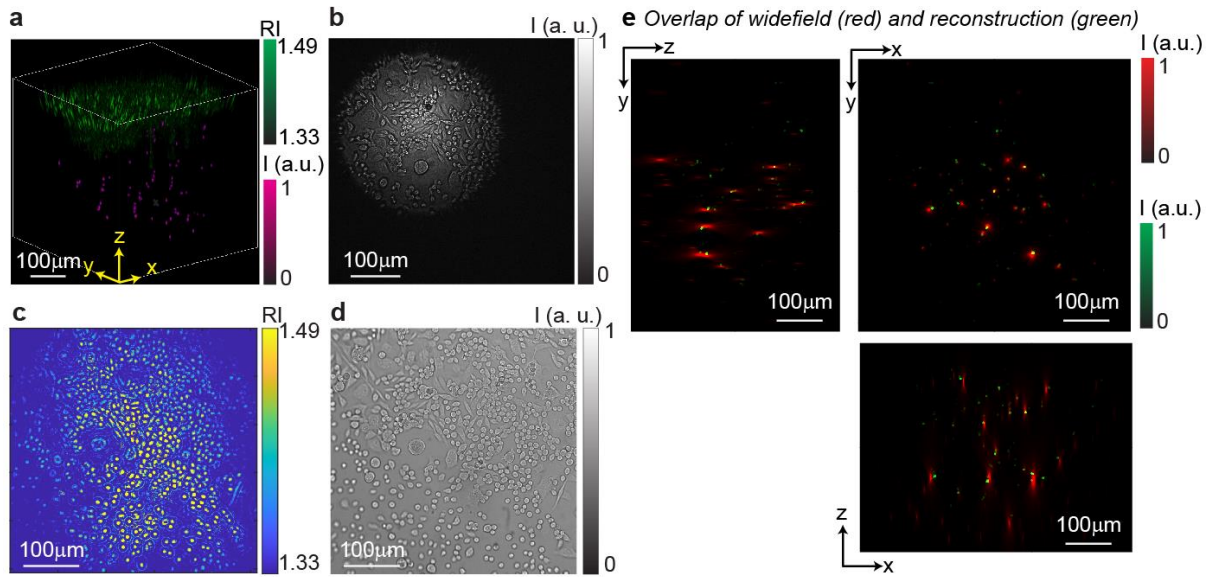

Fig. 3. Experimental results with a sample consisting of fluorescent beads beneath a thin layer of alive CHO cells. **a**. 3D view of fluorescent signals from fluorescent beads (magenta) and RI from CHO cells (green). **b**. A representative raw measurement with one fluorescent bead 'on'. **c**. Reconstructed 3D RI of CHO cells at $\Delta z = 40 \mu m$ below the image plane from 23 forward measurements. **d**. Widefield intensity image of the sample under transmitted infrared illumination to show the ground truth of the cells' lateral positions. **e**. Overlap of the maximum intensity projection (MIP) of the widefield image stack of fluorescence beads excited by 473nm laser (ground truth, red) and the reconstructed 3D distribution of the fluorescence beads' location (green) from 148 forward measurements like **b**. Note this 3D fluorescence image stack is not used in the reconstruction of fluorescence distribution.

In addition, our method also reconstructs the 3D fluorophore locations in the sample from the same raw data. In our case, each fluorescence image only contains one fluorophore, and the 23 reconstructed fluorophores are fairly bright and located relatively deep in the sample. The raw data carries more information about the phase objects because they have higher SNR and the fluorophores in these images illuminate a larger volume compared to the raw data containing dim fluorophores. To find a balance between reconstruction quality and computing burden, we only used these raw data for RI reconstruction. For fluorescence reconstruction, to also reconstruct dim fluorophores and fluorophores that are near the image plane, we select another 116 images (139 in total, with the 23 images used in the RI reconstruction) from the raw data that we collected in the same scanning process as described above. The reconstruction results (Fig. 3f) indicate the location of fluorophores, with intensity being proportional to the sum of fluorescence intensity in the measurement. For validation of our method, we also take a 3D axial-scanned image stack of fluorophores under widefield illumination with a 473nm laser (Fig. 3e). This image stack is used as ground truth for fluorophore positions, and not used in the reconstruction of Fig. 3f. Given the scattering from the CHO cells is weak, the image stack is close to the ground truth of the fluorophores. Compared to localization of fluorophores from the 3D image stack (Fig. 3g), our method can reconstruct the location of most bright fluorophores accurately at high spatial precision (SSIM=0.9786). Since the RI and fluorescence are collected in the same experiment with a single camera, they are automatically registered in the 3D volume (Fig. 3a), unlike previous methods[29] that collect fluorescence and RI information with two different cameras.

Knowing both the fluorescence and the 3D RI information not only provides a multimodal reconstruction of the sample's structural and functional maps, but also can be used to digitally correct multiple-scattering effects in the fluorescence images. To demonstrate, we use a negative fluorescent USAF target as the fluorescence object (instead of fluorescent beads), placed below an opaque glass-bead sample as the scattering phantom, with immersion oil (RI=1.515) in between (Fig. 4a). We first focus on the USAF target through the scattering glass-bead sample and take a widefield fluorescence image in focus (Fig. 4d). Then we move the focal plane above the glass-bead sample and collect scattered fluorescence images as the measurements for RI reconstruction, like in the previous experiments. We reconstruct the 3D RI of the glass-bead sample from 10

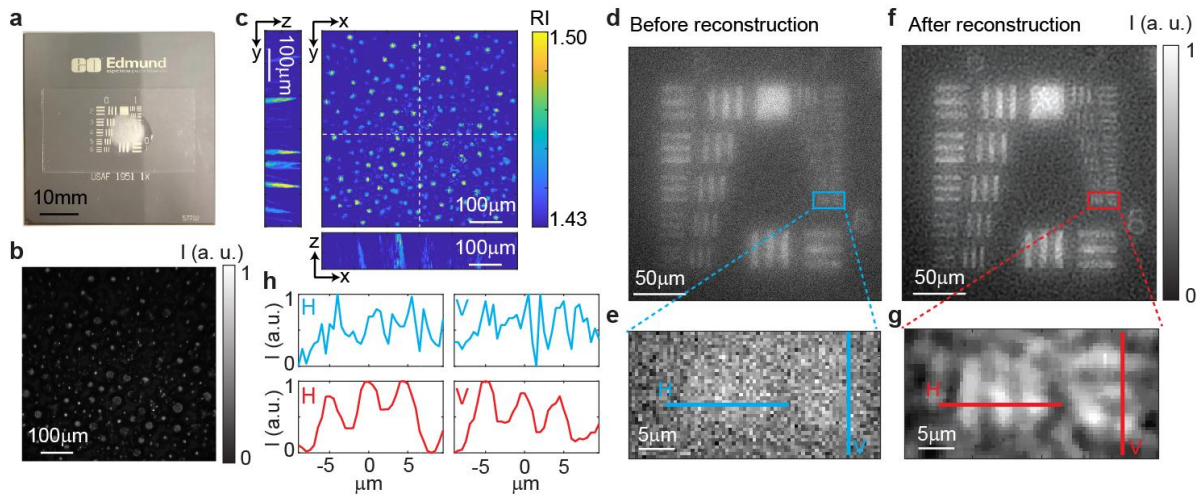

Fig. 4. Multi-modal microscopy for digital correction of multiple scattering in fluorescence images. **A**. A negative fluorescence USAF target is used as the fluorescent sample and a highly-scattering glass-beads phantom on top of it acts as the scattering media. **b.** A representative image of the raw measurements. The glass beads are illuminated by fluorescence from the USAF target instead of fluorescence beads in the previous experiments. **c.** Orthogonal slice views of the reconstructed 3D RI. **d**. Raw fluorescence image of the USAF target under wide-field blue laser illumination. This image is taken in focus, whereas (b) is captured with the system focused at the top surface of the sample. **e**. Zoom-in view of the area in the blue box in (d), containing line pairs of element 6, group 7. **f**. Reconstructed image after correcting multiple scattering with the multi-slice model. **g**. Zoom-in view of the area in the red box in (f). **h**. Normalized intensity profiles along the horizontal direction (H) and the vertical direction (V) in (e, blue) and (g, red).

measurements (Fig. 4b). One representative plane of the reconstructed 3D RI is shown in Fig. 4c.

To digitally correct multiple-scattering effects in the fluorescence image (Fig. 4d-e), we take two steps: we first calculate the scattered PSF by the convolution of a 2D delta function and the reconstructed 3D RI of the scattering phantom (Fig. 4c), assuming the PSF is spatially invariant; we next perform Richardson-Lucy deconvolution on the scattered image (Fig. 4d) with the scattered PSF. The reconstructed image can clearly distinguish the finest line pairs (Element 6 of Group 7, 228.1 line pairs/mm) on the USAF target (Fig. 4f-g), whereas the scattered image cannot. Fig. 4h-i quantitatively compares the normalized intensity profiles of the line pairs in the scattered image (Fig. 4e) and in the reconstructed image (Fig. 4g) in both horizontal (H) and vertical (V) directions. Hence, we have shown that the reconstructed RI can be used to digitally correct multiple-scattering effects and improve image SNR of fluorescence imaging. Our advantage over the previous work[1-16] on imaging through scattering is that we can potentially correct scattering effect for fluorescence objects at any location in 3D inside of the scattering tissue without additional experimental measurements, since we already reconstructed the 3D RI of the whole volume.

In the study, we demonstrated and validated BRIEF, a new imaging method that operates in epi mode and reconstructs 3D fluorescence and 3D RI from only fluorescence images by solving the inverse problem of multiple scattering based on a multi-slice model. We experimentally demonstrate the 3D reconstructed RI of glass beads and alive CHO cells registered with fluorescence beads in the same sample. We also demonstrate an application of BRIEF with the reconstructed RI by digitally correcting multiple scattering effects and improving the SNR of fluorescence images that are taken through a visually opaque phantom.

Our technique not only works for sparse fluorophores but also for dense fluorophores. In the above experiments, we excited only one fluorophore for each measurement, given that the fluorophores are sparsely distributed. For dense fluorophores, we probably cannot excite a single fluorophore at a time but excite multiple fluorophores (say N fluorophores) simultaneously at each measurement (Fig. S1). In this case, since the measured fluorescence image is the sum of the intensity of the scattered light illuminated by each fluorophore, each fluorophore needs to be measured at least N times combined with different fluorophores to acquire mutually independent measurements. As an example, we perform a simulation that contains 10 fluorophores illuminating a 3D phantom ($\Delta RI=0.01$) from below (Fig. S1a). The first result is reconstructed from 10 measurements and each measurement is illuminated by a single fluorophore (Fig. S1b); the second result is also reconstructed from 10 measurements, but each measurement is illuminated by 5 fluorophores simultaneously (Fig. S1c). The result from single fluorophore illumination is better than the result from multiple fluorophores illumination given the same number of measurements ($SSIM_{single}=0.9999$, $SSIM_{multiple}=0.9998$). Therefore, we choose to illuminate a single fluorophore in each measurement in all the experiments above to achieve better reconstruction results, but our technique also works if illuminating multiple fluorophores in each measurement.

The resolution of our method ultimately is limited by the finite numerical aperture, so the axial resolution is poorer than the lateral resolution. Like ODT, the more measurements we take, the higher axial resolution we can achieve. However, more measurements and finer axial grid in the multi-slice model cause heavier computational burden. With prior knowledge of the sparse phase objects, we are able to resolve individual glass beads and cells with dozens of intensity measurements. We could potentially achieve higher axial resolution with more measurements and more slices in $z$.

When imaging through high scattering tissue, our current method may have poor performance due to model mismatching. We assume weak and isotropic scattering in the current model, so we can accurately estimate the 3D position of fluorophores and then use this information to reconstruct 3D RI. However, under strong scattering, 3D localization of fluorophores from the raw measurements will be challenging. Also, highly scattering tissue contains more heterogeneous structures. To reconstruct RI at higher resolution in order to distinguish these structures, our method will require finer grid in the multi-slice model and more raw measurements, which will add more computational burden as we discussed above.

In conclusion, we provide a proof-of-concept experiment of BRIEF to reconstruct RI information from fluorescence images, which is beyond the conventional applications of fluorescence microscopy. BRIEF is a versatile technique and is compatible with both one-photon and multiphoton microscopy, which will facilitate a wide range of applications in biology.

**Methods**

**Data availability.** The datasets generated during and/or analyzed during the current study are available upon reasonable request to the corresponding authors.

**Code availability.** The codes generated during and/or analyzed during the current study are available upon reasonable request to the corresponding authors.

**Computational reconstruction framework**

1. Reconstruct 3D fluorescent objects

The first step is to reconstruct 3D fluorescence objects from dozens 2D fluorescence images taken at the same z plane ($z = z_N$), under the conditions of weak scattering and sparse objects. In our experiments, this step is to reconstruct the 3D locations of the fluorophores. The measurements are taken by turning on a "giant pixel" (20x20 pixels) one-by-one on the DMD to scan across the FOV. The excitation light for $l$ th measurement is denoted by $s(\boldsymbol{r} - \boldsymbol{r}_l)$, illuminating the unknown fluorescent object $o_f(\boldsymbol{r})$. The emitted fluorescence, $o_f(\boldsymbol{r}) \cdot s(\boldsymbol{r} - \boldsymbol{r}_l)$, propagates through the detection system and is imaged by a camera. If the scanned area has a fluorophore, the sum intensity of the image is much higher than the background image and the image is denoted by $I_l(\boldsymbol{r}; z_N)$ for $l$ th measurement; otherwise, the scanned area has no fluorophores, and the image is excluded from the reconstruction. Assuming the scattering is weak, this process can be modeled as

$$I_l(\boldsymbol{r}; z_N) = [o_f(\boldsymbol{r}) \cdot s(\boldsymbol{r} - \boldsymbol{r}_l)] \otimes h_f(\boldsymbol{r}; z_N), \qquad (2)$$

where $h_f(\boldsymbol{r}; z_N)$ is the intensity of PSF measured with single fluorescence bead at the same image plane in prior.

Next, we reconstruct the position of $l$th fluorophore in the raw intensity measurement $\{I_l(\boldsymbol{r}; z_N) | l = 1, 2, \dots M\}$ by solving the inverse problems with LASSO regularizer:

$$e_f(o_f) = \underset{o_f(\boldsymbol{r})}{\mathrm{argmin}} \left\| I_l(\boldsymbol{r}; z_N) - \hat{I}_l(\boldsymbol{r}; z_N) \right\|_2^2 + \tau_f \left\| o_f(\boldsymbol{r}) \right\|_1. \qquad (3)$$

We solve this optimization problems with fast iterative shrinkage-thresholding algorithm (FISTA), which is a first-order gradient descent algorithm.

2. RI reconstruction with known fluorophore location

The fluorescence light emitted from a fluorophore is denoted by a spherical wave $E_{0,l}(r) = \frac{\exp(i\frac{2\pi}{\lambda}n_b r)}{r}$ which is the initial electric field of the multi-slice model at $l$th measurement, where $r$ donates the 3D spatial position vector, $\lambda$ is the center wavelength of the fluorophore and $n_b$ denotes the homogenous RI of the surrounding media. As the input electric field $E_{k,l}(r)$ passes through the $k$th ($k = 1, 2, \ldots N$) slice of the multi-slice model, the input electric field will be pointwise multiplied by the corresponding 2D transmittance function $t_k(r) = \exp(i\frac{2\pi}{\lambda}\Delta z(n_k(r) - n_b))$ at the corresponding $z$ depth, where $\Delta z$ denotes the propagation distance between slices and $n_k(r)$ denotes the complex-valued RI at $k$th layer. The output electric field is propagated in free space to the next slice by multiplying the mathematical operator $\mathcal{P}\{\cdot\}$. According to the angular spe (3) propagation method, $\mathcal{P}\{\cdot\} = \mathcal{F}^{-1}\{\exp\left(-i\Delta z\sqrt{\left(\frac{2\pi}{\lambda}\right)^2 - \|\boldsymbol{k}\|^2}\right)\mathcal{F}\{\cdot\}\}$, where $\mathcal{F}\{\cdot\}$ and $\mathcal{F}^{-1}\{\cdot\}$ c (4) 2D Fourier and inverse Fourier transforms, respectively, and $\boldsymbol{k}$ denotes the spatial frequency vector. This process can be recursively written as

$$E_{k,l}(r) = \mathcal{P}\{t_k(r) \cdot E_{k-1,l}(r)\}.$$

The final forward measurement is the intensity of the exit electric field $E_N(r)$ on the top of the multi-slice volume, which is

$$I_l(r; z) = \left|\mathcal{F}^{-1}\left\{p(\boldsymbol{k}; z) \cdot \mathcal{F}\{E_{N,l}(r)\}\right\}\right|^2, \tag{5}$$

where $p(\boldsymbol{k}; z)$ is the pupil function corresponding to the image plane at $z$.

2. Reconstruction algorithm

We use the estimated position of fluorophores $\hat{o}_f(r)$ and the raw intensity measurements $\{I_l(r; z) | l = 1, 2, \ldots M\}$ to estimate the RI by solving the inverse problems with Tikhonov regularizer:

$$\tag{6}$$

$$e_{RI}(n) = \underset{n(\boldsymbol{r})}{\operatorname{argmin}} \sum_{l} \left\| I_l(\boldsymbol{r}; z) - \hat{I}_l(\boldsymbol{r}; z) \right\|_2^2 + \tau_{RI} \| n(\boldsymbol{r}) \|_2^2.$$

We solve both optimization problems with FISTA as well.

**Optical setup**

The laser source to excite fluorescence objects is a Diode Pumped Solid State (DPSS) laser diode at 473nm wavelength (MBL-N-473A-1W, Ultralasers, Inc., Canada). Current supplies are externally driven by an analog modulation voltage. The laser beam size is expanded to fill the DMD (DLP9000X and the controller V4390, ViALUX, Germany). The DMD is at the relayed image plane, mounted on a rotation base to maximize the output laser power from 'ON' pixels and minimize the diffraction pattern from DMD pitches. Then the patterned beam passes through a tube lens (f=180mm) and the objective lens (XLUMPlanFL N, 20x, NA 1.0, Olympus) to generate illumination spot. The objective lens is mounted on a motorized z-stage (FG-BOBZ-M, Sutter Instrument, U.S.). A customized dichroic mirror (zt473/589/635rpc-UF2, Chroma, U.S.) is placed before the objective lens to reflect the stimulation laser beams while transmitting fluorescence photons emitted from the sample. Fluorescence images are recorded by a camera (Prime95B, Teledyne Photometrix, U.S.) with Micro-Manager software. The samples are placed on a motorized x-y stage (X040528 and the controller MP-285, Sutter Instrument, U.S.). A NI-DAQ (National Instruments, NI PCIe-6363) synthesizes custom analog signals to synchronously modulate the lasers, the DMD, and the camera. A custom MATLAB (MathWorks, U.S.) graphic user interface is used to control the NI DAQ boards, calibrate and align the imaging modalities, and for data acquisition. Custom Python codes are used for data processing and reconstruction.

**Sample preparation**

The first step is to make fluorescence samples for experiments in Fig. 2-3. Red fluorescence beads suspension (R700, Thermo Fisher Scientific, MA) is mixed with PDMS (Sylgard 184, Dow Inc, MI) in the ratio of 1:550. Base elastomer and curing agent of PDMS are mixed in the ratio of

10:1. After mixing, we use a vacuum desiccator to remove the air bubbles in the mixture for 30min. Then the mixture is poured onto clean coverslips and heated at 100 Celsius for 35min. A fluorescence sample is finished. For the experiment in Fig. 2, we mix glass beads (SLGMS-2.5, Cospheric LLC, CA) with PDMS and then remove air bubbles with a vacuum desiccator. The mixture is poured onto the fluorescence sample and heated for curing. For the experiment in Fig. 3, we place Chinese Hamster Ovary (CHO) cells on the fluorescence samples coated with poly-D-lysin (Sigma-Aldrich, MO) one day before the experiment. CHO cells were maintained in Han's F-12 medium with L-glutamine (Thermo Fisher Scientific, MA) before the experiment and placed in phosphate-buffered saline (PBS, Thermo Fisher Scientific, MA) during the experiment. The glass-bead sample in Fig. 4 is made by mixing glass beads with PDMS in the ratio of 1:10 and then degasifying the mixture and curing it as described above. The glass-bead sample is place onto a negative fluorescence USAF target (Edmund Optics, CA) with immersion oil (Resolve, Thermo Fisher Scientific, MA) in between.

**Supplement figures**

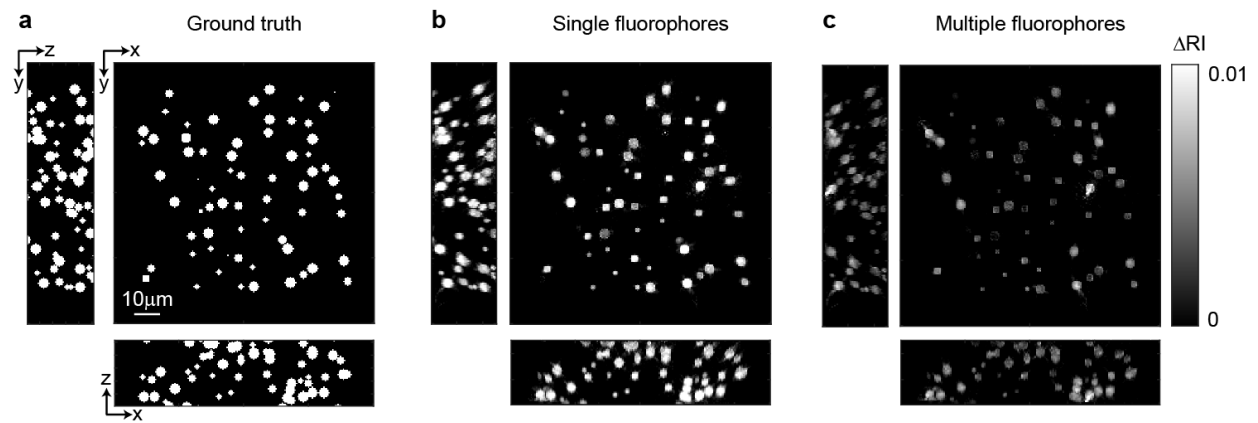

Fig. S1. Simulation results comparing single fluorophore illumination to multiple fluorophores illumination. The sample contains 10 fluorophores below the 3D RI phantom. (a) Ground truth of 3D RI. (b) Reconstructed 3D RI from 10 measurements, each measurement contains only a single fluorophore. (c) Reconstructed 3D RI from 10 measurements, each measurement contains 5 fluorophores that are illuminated simultaneously. We randomly select the 10 combinations of 5 fluorophores from all $C(10,5) = 252$ possible combinations.